\documentclass[prd, nofootinbib, aps, twocolumn, superscriptaddress, showpacs, preprintnumbers, amsmath, amssymb, floatfix]{revtex4}

\usepackage{graphicx}

\newcommand{\nc}{\newcommand}
\nc{\hef}{\ensuremath{^4\mathrm{He}}}
\nc{\het}{\ensuremath{^3\mathrm{He}}}
\nc{\lisx}{\ensuremath{^6\mathrm{Li}}}
\nc{\lisv}{\ensuremath{^7\mathrm{Li}}}
\nc{\bes}{\ensuremath{^7\mathrm{Be}}}
\nc{\beet}{\ensuremath{^8\mathrm{Be}}}
\nc{\ben}{\ensuremath{^9\mathrm{Be}}}
\nc{\dm}{{\rm D}}
\nc{\hefm}{{\rm ^4He}}
\nc{\hetm}{{\rm ^3He}}
\nc{\lisxm}{{\rm ^6Li}}
\nc{\lisvm}{{\rm ^7Li}}
\nc{\besm}{{\rm ^7Be}}
\nc{\beetm}{{\rm ^8Be}}
\nc{\benm}{{\rm ^9Be}}
\nc{\bs}{(N$X^-$)}
\nc{\xm}{$X^-$}
\nc{\xp}{$X^+$}
\nc{\xz}{$X^0$}
\nc{\bex}{(\bes\xm)}
\nc{\bexm}{(\besm X^-)}
\nc{\px}{($p$\xm)}
\nc{\Ox}{\ensuremath{\mathrm{O}}}
\nc{\Fe}{\ensuremath{\mathrm{Fe}}}
\nc{\Hyd}{\ensuremath{\mathrm{H}}}
\nc{\Be}{\ensuremath{\mathrm{Be}}}
\nc{\tauX}{\ensuremath{\tau_{X^-}}}
\nc{\YX}{\ensuremath{Y_{X^-}}}
\nc{\YXdec}{\ensuremath{Y^{\mathrm{dec}}_{X^-}}}
\nc{\Ysldec}{\ensuremath{Y^{\mathrm{dec}}_{{\widetilde{l}_1}}}}
\nc{\Ysldecm}{\ensuremath{Y^{\mathrm{dec}}_{{\widetilde{l}_1^-}}}}
\newcommand{\Order}{{\cal O}}   % e.g. terms up to $\Order(g^2)$
\newcommand{\keV}{\mathrm{keV}}
\newcommand{\MeV}{\mathrm{MeV}}
\newcommand{\GeV}{\mathrm{GeV}}
\newcommand{\TeV}{\mathrm{TeV}}
\newcommand{\Mpc}{\mathrm{Mpc}}
\newcommand{\km}{\mathrm{km}}

\newcommand{\seconds}{\mathrm{s}}

\newcommand{\gravitino}{{\widetilde{G}}}
\newcommand{\axino}{{\widetilde{a}}}
\newcommand{\maxino}{m_{\axino}}
\newcommand{\ax}{\ensuremath{\widetilde{a}}}
\newcommand{\slepton}{{\tilde{l}_1}}
\newcommand{\stauone}{{\widetilde{\tau}_1}}
\newcommand{\stau}{{\widetilde{\tau}}}
\newcommand{\mstau}{m_{\widetilde{\tau}}}

\newcommand{\stauR}{\ensuremath{{\widetilde{\tau}_{\mathrm{R}}}}}

\newcommand{\bino}{{\widetilde B}}

\newcommand{\mbino}{m_{\widetilde{B}}}

\newcommand{\gluino}{{\widetilde g}}
\newcommand{\neutralino}{{\widetilde \chi}^{0}_{1}}

\newcommand{\lepton}{\ensuremath{l}}

\newcommand{\mZ}{\ensuremath{m_{\mathrm{Z}}}}

\newcommand{\quark}{\mathrm{q}}
\newcommand{\antiquark}{\bar{\mathrm{q}}}

\newcommand{\mslepton}{m_{\slepton}}

\newcommand{\PQ}{\mathrm{PQ}}
\newcommand{\CaYY}{C_{\mathrm{aYY}}}

\newcommand{\NTP}{\mathrm{NTP}}
\newcommand{\TP}{\mathrm{TP}}
\newcommand{\thermal}{\mathrm{therm}}

\newcommand{\freezeout}{\mathrm{f}}
\newcommand{\CDM}{\mathrm{dm}}

\newcommand{\OmegaDM}{\Omega_{\mathrm{dm}}}

\newcommand{\Reheating}{\mathrm{R}}
\newcommand{\TR}{T_{\Reheating}}

\newcommand{\TRmax}{T_{\Reheating}^{\max}}

\newcommand{\Color}{\mathrm{c}}
\newcommand{\Weak}{\mathrm{L}}
\newcommand{\Hypercharge}{\mathrm{Y}}
\newcommand{\Lisix}{{}^6 \mathrm{Li}}
\newcommand{\Hefour}{{}^4 \mathrm{He}}

\newcommand{\deuterium}{\mathrm{D}}

\newcommand{\gagg}{g_{a\gamma\gamma}}

\newcommand{\SUSY}{\mathrm{SUSY}}
\newcommand{\tot}{\mathrm{tot}}
\newcommand{\LL}{\mathrm{LL}}

\newcommand{\be}{\begin{equation}}
\newcommand{\ee}{\end{equation}}
\newcommand{\bea}{\begin{eqnarray}}
\newcommand{\eea}{\end{eqnarray}}
\newcommand{\benn}{\begin{displaymath}}
\newcommand{\eenn}{\end{displaymath}}
\newcommand{\beann}{\begin{eqnarray*}}
\newcommand{\eeann}{\end{eqnarray*}}
\begin{document}
% 
% ___ Preprint Numbers ___________________________________________________
%
\preprint{MPP--2009--12, ZU--TH 06/09}
%
%
% ___ Preamble ______________________________________________________
%
\title{Upper Limits on the Peccei--Quinn Scale and on the Reheating Temperature\\
  in Axino Dark Matter Scenarios}
\author{Ayres Freitas}
\email{afreitas@pitt.edu}
\affiliation{Department of Physics and Astronomy, 
University of Pittsburgh, 
PA 15260, USA}
\author{Frank Daniel Steffen}
\email{steffen@mppmu.mpg.de}
\affiliation{Max-Planck-Institut f\"ur Physik, 
F\"ohringer Ring 6,
D--80805 Munich, Germany}
\author{Nurhana Tajuddin} 
\email{nurhana@physik.uzh.ch}
\affiliation{Institut f\"ur Theoretische Physik, 
Universit\"at Z\"urich, 
Winterthurerstrasse 190, 
CH--8057~Z\"urich, Switzerland}
\author{Daniel Wyler} 
\email{wyler@physik.unizh.ch}
\affiliation{Institut f\"ur Theoretische Physik, 
Universit\"at Z\"urich, 
Winterthurerstrasse 190, 
CH--8057~Z\"urich, Switzerland}
%
%
%\date{\today}
%
% ___ Abstract _________________________________________________________
%
\begin{abstract}
  Considering axino cold dark matter scenarios with a long-lived
  charged slepton, we study constraints on the Peccei--Quinn scale
  $f_a$ and on the reheating temperature $\TR$
  imposed by the dark matter density and by big bang nucleosynthesis
  (BBN).
  For an axino mass compatible with large-scale structure,
  $\maxino\gtrsim 100~\keV$, temperatures above $10^9\,\GeV$ become
  viable for $f_a>3\times 10^{12}\,\GeV$.
  We calculate the slepton lifetime in hadronic axion models.
  With the dominant decay mode being two-loop suppressed, this
  lifetime can be sufficiently large to allow for primordial bound
  states leading to catalyzed BBN of lithium--6 and beryllium--9.
  This implies new upper limits on $f_a$ and on $\TR$ that depend on
  quantities which will be probed at the Large Hadron Collider.
\end{abstract}
%
% 04.65.+e      Supergravity (see also 12.60.Jv Supersymmetric models)
%% 12.60.Jv     Supersymmetric models (see also 04.65.+e Supergravity)
% 14.80.Ly      Supersymmetric partners of known particles
%% 95.30.Cq     Elementary particle processes
%% 95.35.+d     Dark matter
%% 98.80.Cq     Particle-theory and field-theory models of the early Universe
%
%\keywords{}
\maketitle
%
% __________________________________________________________________
\section{Introduction}
% __________________________________________________________________

In supersymmetric (SUSY) extensions of the Standard Model with
conserved R-parity, the lightest supersymmetric particle (LSP) is
stable and thus a compelling dark matter candidate.
While the lightest neutralino $\neutralino$ or the gravitino
$\gravitino$ are often considered to be the LSP, the axino $\axino$ is
also a well-motivated LSP candidate and hence an equally compelling
dark matter
candidate~\cite{Bonometto:1993fx,Covi:1999ty,Covi:2001nw,Covi:2004rb,Brandenburg:2004du,Steffen:2008qp,Baer:2008yd}
beyond the minimal supersymmetric Standard Model (MSSM).

The axino $\axino$ is the fermionic partner of the axion in SUSY
extensions of the Standard Model in which the Peccei--Quinn (PQ)
mechanism is embedded as a solution of the strong CP problem.  Because
its interactions are suppressed by the PQ scale $f_a\gtrsim 6\times
10^8\,\GeV$~\cite{Amsler:2008zz,Sikivie:2006ni,Raffelt:2006rj,Raffelt:2006cw},
the axino can be classified as an extremely weakly interacting
particle (EWIP). With the axino being the LSP, the lightest Standard
Model superpartner or lightest ordinary superpartner (LOSP) is
unstable and can thus be an electrically charged particle such as a
charged slepton $\slepton$.  For example, the lighter stau
$\stauone$---which is the superpartner of the tau lepton $\tau$---is
the LOSP in a large part of the parameter space of the constrained
MSSM (CMSSM).  Due to the extremely suppressed axino interaction
strength, such an LOSP would be long-lived and would appear as a
quasi-stable charged particle in the collider detectors.  Its ultimate
decay into the $\axino$ LSP will often occur outside of those
detectors. Some decays however may be accessible experimentally and
may allow one to probe the PQ scale at
colliders~\cite{Brandenburg:2005he}. While an axino LSP
identification~\cite{Brandenburg:2005he} will require challenging
experimental setups~\cite{Hamaguchi:2006vu}, quasi-stable $\slepton$'s
can appear as a first hint for the existence of SUSY and of the axino
LSP at the Large Hadron Collider (LHC) already within the next three
years.

In this Letter we focus on the axino LSP case with a long-lived
$\slepton$ LOSP and in particular on scenarios in which the axino
provides the dominant contribution to the dark matter
density~\cite{Spergel:2006hy}
\begin{equation}
        \Omega_{\CDM}^{3\sigma}h^2=0.105^{+0.021}_{-0.030} 
%        \ ,
\label{Eq:OmegaDM}
\end{equation} 
with $h=0.73^{+0.04}_{-0.03}$ denoting the Hubble constant in units of
$100~\km\,\Mpc^{-1}\seconds^{-1}$.
The $3\sigma$ range indicated rests on a representative six-parameter
``vanilla'' mod\-el.
 
The thermally produced (TP) axino density $\Omega_{\axino}^{\TP}$ must
not exceed $\OmegaDM$. This puts upper limits on the post-inflationary
reheating temperature
$\TR$~\cite{Covi:2001nw,Brandenburg:2004du,Choi:2007rh,Kawasaki:2007mk,Baer:2008yd}.
These $\TR$ limits---which depend on the axino mass $m_{\ax}$ and on
the PQ scale $f_a$---can be very restrictive for models of inflation
and of baryogenesis. For example, $\TR\lesssim 10^6\,\GeV$ is found
for $f_a=10^{11}~\GeV$ and
$m_{\ax}=100~\keV$~\cite{Brandenburg:2004du}.  Indeed, for
$m_{\ax}\gtrsim 100~\keV$, temperatures above $10^9\,\GeV$ can become
viable only for larger values of the PQ scale, $f_a\gtrsim 3\times
10^{12}~\GeV$, if a standard thermal history is assumed.%
\footnote{Depending on the model, the saxion---which is the bosonic
  partner of the axino that appears in addition to the axion---can be
  a late decaying particle and as such be associated with significant
  entropy
  production~\cite{Kim:1992eu,Lyth:1993zw,Chang:1996ih,Hashimoto:1998ua}.
  This could affect cosmological constraints~\cite{Kawasaki:2007mk}
  including those considered in this work. Leaving a study of saxion
  effects for future work, we assume in this Letter a standard thermal
  history and thereby that those effects are negligible.}
While $\TR\gtrsim 10^9\,\GeV$ is required, e.g., by standard thermal
leptogenesis with hierarchical right-handed
neutrinos~\cite{Fukugita:1986hr,Davidson:2002qv,Buchmuller:2004nz,Blanchet:2006be,Antusch:2006gy},
we show in this work that $f_a\gtrsim 3\times 10^{12}~\GeV$ can be
associated with restrictive BBN constraints due to the long-lived
$\slepton$ LOSP
and its potential to form primordial bound states.
In fact, we find that those BBN constraints imply upper limits on
$f_a$ and thereby new upper limits on $\TR$.

We consider hadronic (or KSVZ) axion
models~\cite{Kim:1979if,Shifman:1979if} in a SUSY
setting~\cite{Kim:1983ia}. In this class of models, the axino couples
to the MSSM particles only indirectly through loops of heavy KSVZ
(s)quarks. Thereby, the dominant 2-body decay of the $\slepton$ LOSP
into the associated lepton and the axino is described in leading order
by 2-loop diagrams~\cite{Covi:2004rb,Brandenburg:2005he}.  Using a
heavy mass expansion, we evaluate the 2-loop diagrams explicitly and
obtain the decay width that governs the $\slepton$ lifetime
$\tau_{\slepton}$.
For a given thermal freeze-out yield of negatively charged
$\slepton^-$'s, $Y_{\slepton^-}$, our $\tau_{\slepton}$ result allows
us to infer the BBN constraints associated with primordial $\lisxm$
and $\ben$ production that can be catalyzed by
$\slepton^-$-nucleus-bound-state
formation~\cite{Pospelov:2006sc,Pospelov:2007js,Pospelov:2008ta}.
While BBN constraints were often assumed to play only a minor role in
the axino LSP case, we explore the ones from bound-state effects
explicitly and show that they impose new restrictive limits on $f_a$
and $\TR$.

Before proceeding, let us comment on axion physics. We assume a
cosmological scenario in which the spontaneous breaking of the PQ
symmetry occurs before inflation leading to $\TR<f_a$ so that no PQ
symmetry restoration takes place during inflation or in the course of
reheating. Since axions are never in thermal equilibrium for the large
$f_a$ values considered, their relic density $\Omega_a$ is governed by
the initial misalignment angle $\Theta_i$ of the axion field with
respect to the CP-conserving position;
cf.~\cite{Beltran:2006sq,Sikivie:2006ni,Steffen:2008qp} and references
therein.  With a sufficiently small $\Theta_i$ being an option,
$\Omega_a\ll\OmegaDM$ is possible even for $f_a$ as large as
$10^{14}\,\GeV$. 
We assume $\Omega_a\ll\OmegaDM$
to keep the presented constraints conservative.

The remainder of this Letter is organized as follows. 
In the next section we review the upper limits on $\TR$ imposed by
$\Omega_{\axino}^{\TP}\leq\OmegaDM$ which provide our motivation to
consider $f_a\gtrsim 3\times 10^{12}~\GeV$.
Section~\ref{Sec:ChargedSleptonLOSP} presents the results for the
$\slepton$ NLSP lifetime obtained from our 2-loop calculation.
Section~\ref{Sec:CBBNaxino} explores the BBN constraints from
$\slepton$-nucleus-bound-state formation.
In Sect.~\ref{Sec:ProbingTR}, we show that those BBN constraints imply
new $\TR$ limits if the considered axino LSP scenario is realized in
nature.
Analytic expressions that approximate the obtained limits in a
conservative way are derived in Sect.~\ref{Sec:Discussion}.

% __________________________________________________________________
\section{Constraints on \boldmath$\TR$}
% __________________________________________________________________

Because of their extremely weak interactions, the temperature
$T_{\freezeout}$ at which axinos decouple from the thermal plasma in
the early Universe can be very high, e.g., $T_{\freezeout} \gtrsim
10^9\,\GeV$ for $f_a\gtrsim
10^{11}\,\GeV$~\cite{Rajagopal:1990yx,Brandenburg:2004du} or
$T_{\freezeout} \gtrsim 10^{11}\,\GeV$ for $f_a\gtrsim
10^{12}\,\GeV$~\cite{Brandenburg:2004du}.
Accordingly, axinos decouple as a relativistic species in scenarios
with $\TR>T_{\freezeout}$. The resulting relic density is then
insensitive to the precise value of $\TR$~\cite{Rajagopal:1990yx}:
$\Omega_{\ax}^{\thermal}h^2\simeq m_{\axino}/(2~\keV)$.
Moreover, $\Omega_{\ax}^{\thermal}\leq \OmegaDM$ implies
$m_{\ax}\lesssim 0.2~\keV$. For a scenario with
$\Omega_{\ax}^{\thermal}\simeq\OmegaDM$, this is in conflict with
large-scale structure which requires a smaller present free-streaming
velocity of axino dark matter and thereby $m_{\ax}\gtrsim 1~\keV$;
cf.\ Sect.~5.2 and Table~1 of Ref.~\cite{Steffen:2006hw}.
Focussing on scenarios in which axinos provide the dominant component
of cold dark matter with a negligible present free-streaming velocity,
$\maxino\gtrsim 100~\keV$, we thus assume $\TR<T_{\freezeout}$ in the
remainder of this work.

In scenarios with $\TR<T_{\freezeout}$, axino dark matter can be
produced efficiently in scattering processes of particles that are in
thermal equilibrium within the hot MSSM
plasma~\cite{Asaka:2000ew,Covi:2001nw,Brandenburg:2004du,Gomez:2008js}.
The efficiency of this thermal axino production is sensitive to $\TR$
and $f_a$ and the associated relic density
reads~\cite{Brandenburg:2004du}%
\footnote{We refer to $\TR$ as the initial temperature of the
  radiation-dominated epoch. Relations to $\TR$ definitions in terms
  of the decay width of the inflaton field can be established in the
  way presented explicitly for the $\gravitino$ LSP case in
  Ref.~\cite{Pradler:2006hh}.}
\bea
        \Omega_{\axino}^{\TP}h^2
        &\simeq&
        5.5\,g_\mathrm{s}^6(\TR) \ln\left(\frac{1.211}{g_\mathrm{s}(\TR)}\right) 
        \left(\frac{10^{11}\,\GeV}{f_a}\right)^{\! 2}\!\!
\nonumber\\
        &&
        \times
        \bigg(\frac{m_{\ax}}{0.1~\MeV}\bigg)\!
        \left(\frac{T_R}{10^7\,\GeV}\right)
%        \ ,
\label{Eq:AxinoDensityTP}
\eea
with the strong coupling $g_\mathrm{s}$ and the axion-model-dependent
color anomaly of the PQ symmetry absorbed into $f_a$.%
\footnote{For the hadronic axion models considered below, the color
  anomaly is $N=1$ so that~(\ref{Eq:AxinoDensityTP}) applies directly,
  i.e., without the need to absorb $N$ into the definition of $f_a$.}
Using hard thermal loop (HTL) resummation together with the
Braaten-Yuan prescription~\cite{Braaten:1991dd}, this expression has
been derived within SUSY QCD in a consistent gauge-invariant treatment
that requires weak couplings $g_\mathrm{s}(\TR)\ll 1$ and thus high
temperatures. Accordingly, (\ref{Eq:AxinoDensityTP}) is most reliable
for $T\gg 10^4\,\GeV$~\cite{Brandenburg:2004du}.%
\footnote{For thermal axino production at lower temperatures,
  cf.~\cite{Gomez:2008js}.}
Note that we evaluate $g_s(\TR)=\sqrt{4\pi\alpha_s(\TR)}$ according to
its 1-loop renormalization group running within the MSSM from
$\alpha_s(\mZ)=0.1176$ at $\mZ=91.1876~\GeV$.

\begin{figure}[t!]
\includegraphics*[width=.495\textwidth]{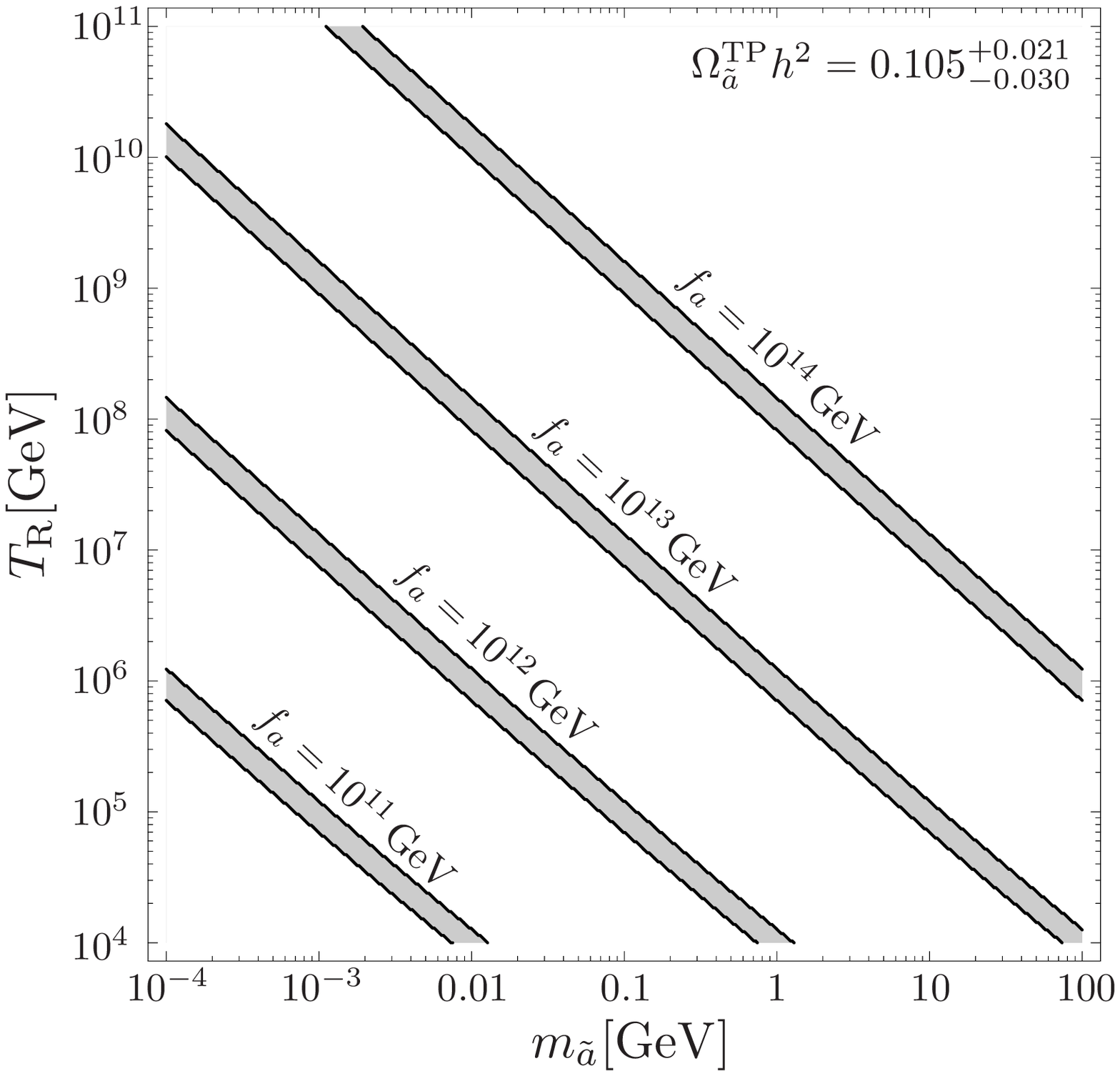}
\caption{Upper limits on the reheating temperature $\TR$ as a function
  of the axino mass $\maxino$ in scenarios with axino cold dark matter
  for $f_a=10^{11}$, $10^{12}$, $10^{13}$, and $10^{14}\,\GeV$ (as
  labeled). For $(m_{\ax}, \TR)$ combinations within the gray bands,
  the thermally produced axino density $\Omega_{\axino}^{\TP}h^2$ is
  within the nominal $3\sigma$ range~(\ref{Eq:OmegaDM}). For given
  $f_a$, the region above the associated band is disfavored by
  $\Omega_{\ax}^{\TP}h^2>0.126$.}
\label{Fig:TR_Limits}
\end{figure}

In Fig.~\ref{Fig:TR_Limits}, 
%
% __________________________________________________________________
% __________________________________________________________________
%
$(m_{\ax}, \TR)$ regions in which the thermally produced axino
density~(\ref{Eq:AxinoDensityTP}) is within the nominal $3\sigma$
range~(\ref{Eq:OmegaDM}) are indicated for $f_a$ values between
$10^{11}\,\GeV$ and $10^{14}\,\GeV$ by gray bands (as labeled).
For given values of $m_{\ax}$ and $f_a$, $\TR$ values above the
corresponding band are disfavored by $\Omega_{\ax}^{\TP}>\OmegaDM$;
see
also~\cite{Covi:2001nw,Brandenburg:2004du,Choi:2007rh,Kawasaki:2007mk,Baer:2008yd}.
From (\ref{Eq:AxinoDensityTP}) and Fig.~\ref{Fig:TR_Limits}, one can
see that the viability of temperatures above $10^9\,\GeV$ points to
$f_a>3\times 10^{12}\,\GeV$ if one insists on cold axino dark matter,
$\maxino\gtrsim 100~\keV$, providing the dominant component of
$\OmegaDM$.
Those $f_a$ values and $\maxino\lesssim 1~\GeV$ are thereby favored by
the viability of standard thermal leptogenesis with hierarchical
right-handed
neutrinos~\cite{Fukugita:1986hr,Davidson:2002qv,Buchmuller:2004nz,Blanchet:2006be,Antusch:2006gy}.

% __________________________________________________________________
\section{The Charged Slepton LOSP Case}
\label{Sec:ChargedSleptonLOSP}
% __________________________________________________________________

While the $\TR$ limits discussed above are independent of the LOSP, we
turn now to the phenomenologically attractive case in which the LOSP
is a charged slepton $\slepton$. To be specific, we focus on the
$\stauone$ LOSP case under the simplifying assumption that the lighter
stau is purely `right-handed,' $\stauone=\stauR$, which is a good
approximation at least for small $\tan\beta$. The
$\neutralino$--$\stauone$ coupling is then dominated by the bino
coupling. For further simplicity, we also assume that the lightest
neutralino is a pure bino: $\neutralino=\bino$.

We consider SUSY hadronic axion models in which the interaction of the
axion multiplet $\Phi$ with the heavy KSVZ quark multiplets $Q_1$ and
$Q_2$ is described by the superpotential
\begin{equation}
        W_{\PQ}=y\Phi Q_1 Q_2 
\label{Eq:SuperpotentialPQ}
\end{equation} 
with the quantum numbers given in Table~\ref{Tab:Q_quantum_numbers}
%
%
% __________________________________________________________________
\begin{table}[t]
    \caption{The axion multiplet $\Phi$, the heavy KSVZ quark multiplets 
    $Q_{1,2}$, and the associated quantum numbers considered in this work.}
\label{Tab:Q_quantum_numbers}
\begin{center}
  \renewcommand{\arraystretch}{1.25}
\begin{tabular*}{3.25in}{@{\extracolsep\fill}rcl}
\hline
chiral multiplet           
& U(1)$_{\PQ}$ 
& (SU(3)$_\Color$,\,SU(2)$_\Weak$)$_{\Hypercharge}$
\\ \hline
$\Phi\,\,=\,\,\phi\,\,+\,\sqrt{2}\chi\theta+F_{\Phi}\theta\theta$                
& +1 
& ($\bf{1}$,\,$\bf{1}$)$_0$
\\
$Q_1=\widetilde{Q}_1+\sqrt{2}q_1\theta+F_1\theta\theta$                 
& -1/2 
& ($\bf{3}$,\,$\bf{1}$)$_{+e_Q}$
\\
$Q_2=\widetilde{Q}_2+\sqrt{2}q_2\theta+F_2\theta\theta$                 
& -1/2 
& ($\bf{3^*}$,\,$\bf{1}$)$_{-e_Q}$
\\
\hline
\end{tabular*}
\end{center}
\end{table}
% __________________________________________________________________
%
and the Yukawa coupling $y$.
From the 2-component fields of Table~\ref{Tab:Q_quantum_numbers}, the
4-component fields describing the axino and the heavy KSVZ quark are
given, respectively, by
\begin{equation}
        \axino = \begin{pmatrix}\chi \\ \bar{\chi}\end{pmatrix}
        \quad \mbox{and} \quad
        Q =  \begin{pmatrix} q_1 \\ \bar{q}_2 \end{pmatrix}
\ .
\end{equation}
For the heavy KSVZ (s)quark masses, we use the SUSY limit
$M_{\widetilde{Q}_{1,2}}=M_Q=y\langle\phi\rangle=y f_a / \sqrt{2}$
with both $y$ and $f_a$ taken to be real by field redefinitions.
The phenomenological constraint $f_a\gtrsim 6\times
10^8\,\GeV$~\cite{Amsler:2008zz,Sikivie:2006ni,Raffelt:2006rj,Raffelt:2006cw}
thus implies a large mass hierarchy between the KSVZ (s)quarks and the
weak and the soft SUSY mass scales for $y=\Order(1)$,
\begin{equation}
M_{\widetilde{Q}_{1,2}}, M_Q\gg \mZ, m_{\SUSY} 
\ .
\label{Eq:MassHierarchy}
\end{equation}

Before proceeding, let us recall axion and axino interactions to
clarify the definition of $f_a=\sqrt{2}\langle\phi\rangle$ in the
considered models.
By integrating out the heavy KSVZ (s)quarks, axion-gluon and
axion-photon interactions are obtained as described by the effective
Lagrangians
\begin{eqnarray}
   {\cal L}_{a gg} 
   = 
   \frac{g_\mathrm{s}^2}{32\pi^2f_a}\,
   a\,G_{\mu\nu}^a\widetilde{G}^{a\mu\nu}
\label{Eq:Laxiongluon}\\
   {\cal L}_{a \gamma \gamma} 
   = 
   \frac{e^2 C_{a \gamma \gamma}}{32\pi^2f_a}\,
   a\,F_{\mu\nu}\widetilde{F}^{\mu\nu}\ ,
\label{Eq:Laxionphoton}
\end{eqnarray}
where $G_{\mu\nu}^a$ and $F_{\mu\nu}$ are the gluon and
electromagnetic field strength tensors, respectively, whose duals are
given by
$\widetilde{G}^{a}_{\mu\nu}=\epsilon_{\mu\nu\rho\sigma}G^{a\:\!\rho\sigma}/2$
and
$\widetilde{F}_{\mu\nu}=\epsilon_{\mu\nu\rho\sigma}F^{\rho\sigma}/2$;
$e^2=4\pi \alpha$.
After chiral symmetry breaking,
\begin{equation}
  C_{a\gamma\gamma}=6 e_Q^2 - \frac{2}{3}\frac{4+z}{1+z}
\label{Eq:Cagammagamma}
\end{equation}
for the models described by~(\ref{Eq:SuperpotentialPQ}) and
Table~\ref{Tab:Q_quantum_numbers}, where $z=m_u/m_d\simeq 0.56$
denotes the ratio of the up and down quark masses.
The corresponding interactions of axinos with gluons and gluinos
$\gluino$ are obtained as described by 
\begin{equation}
  {\cal L}_{\axino\gluino g}
  = 
  i\,\frac{g_\mathrm{s}^2}{64\pi^2 f_a}\,
  \bar{\axino}\,\gamma_5
  \left[\gamma^{\mu},\gamma^{\nu}\right]\,
  \gluino^a\, G^a_{\mu\nu}
%  \ ,
\label{Eq:L_agG}
\end{equation}
and as used in the derivation of~(\ref{Eq:AxinoDensityTP}).

\begin{figure}[b!]
\includegraphics*[width=.475\textwidth]{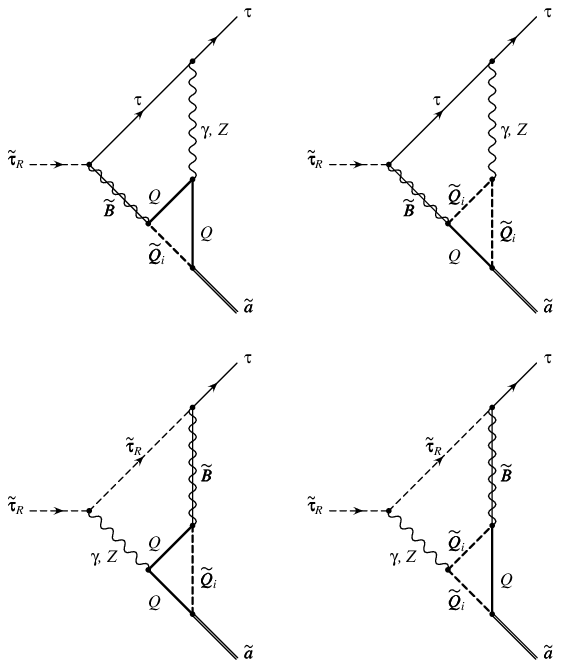}
\caption{Feynman diagrams of the dominant contributions to the
  stau NLSP decay $\stauR\to\tau\axino$ in a SUSY hadronic axion
  model with one KSVZ quark $Q=(q_1,\bar{q}_2)^{\mathrm{T}}$ and the
  associated squarks $\widetilde{Q}_{1,2}$. The considered quantum
  numbers are given in Table~\ref{Tab:Q_quantum_numbers}.  For
  simplicity, the lightest neutralino is assumed to be a pure bino
  $\neutralino=\bino$ and the tau mass is neglected.}
\label{Fig:Stau_Two-Body}
\end{figure}

In R-parity conserving settings in which the $\stauR$ LOSP is the
NLSP, its lifetime $\tau_{\stau}$ is governed by the decay
$\stauR\to\tau\axino$. 
For the models given by~(\ref{Eq:SuperpotentialPQ})
and Table~\ref{Tab:Q_quantum_numbers}, the Feynman diagrams of the
dominant contributions to the 2-body stau NLSP decay
$\stauR\to\tau\axino$ are shown in Fig.~\ref{Fig:Stau_Two-Body}.
%
% __________________________________________________________________
% __________________________________________________________________
%
Since $m_\tau \ll \mstau$, we work in the limit $m_{\tau}\to 0$.
The decay amplitude depends on the parameters of the heavy (s)quark
sector through their masses $M_Q=y f_a/\sqrt{2}$, the Yukawa coupling
$y$, and the gauge couplings $e e_Q$.
In fact, in the calculation of the 2-loop diagrams, the
hierarchy~(\ref{Eq:MassHierarchy}) allows us to make use of a heavy
mass expansion in powers of $1/f_a$~\cite{Schilling:2005dr}. In this
asymptotic expansion, it is sufficient to calculate the leading term
of the amplitude $\propto 1/f_a$ since the sub-leading terms ($\propto
1/f_a^2$) are suppressed by many orders of magnitude. Details of this
calculation and the full result of the leading term will be presented
in a forthcoming publication~\cite{Freitas:2009xxx}.  The dominant
leading logarithmic (LL) part of the partial width is given by
\bea
  && \!\!\!\!\!\!\!\!\!\!\!\!\!\!\!\!\!\!
  \Gamma_{\tot}^{\stauR} 
  \approx 
  \Gamma(\stauR\to\tau\axino)_{\LL} 
\\
  && \!\!\!\!\!\!\!\!\!\!\!\!\!\!\!\!
  = 
  \frac{81\,\alpha^4 e_Q^4}{128\pi^5\cos^8 \theta_{W}}
  \, 
  \frac{\mstau\,\mbino^2}{f_a^2}\! 
%  \, 
  \left(\!1-\frac{\maxino^2}{\mstau^2}\right)^{\!\!2} 
  \ln^2 \!\left(\frac{y f_a}{\sqrt{2}\mstau}\right)\!,
\label{Eq:GammaLL}
\eea
where $\alpha$ denotes the fine structure constant, $\mbino$ the bino
mass, and $\theta_{W}$ the weak mixing angle.%
\footnote{We use
  $\alpha=\alpha^{\overline{\mathrm{MS}}}(\mZ)=1/129$~\cite{Jegerlehner:2008sd}
   and $\sin^2\theta_{W}=1-m_W^2/m_Z^2=0.2221$.}
However, all numerical results shown in the plots below
rest on the full calculation.%
\footnote{Note that the 3-body decay $\stauR\to\tau\axino\gamma$
  occurs already at the 1-loop level. The corresponding amplitude
  however is not enhanced by $\ln(y f_a/\sqrt{2}\mstau)$ which can be
  as large as 20.4--27.3 for $\mstau/y=100~\GeV$ and
  $f_a=10^{11}$--$10^{14}\,\GeV$. In fact, the branching ratio of
  $\stauR\to\tau\axino\gamma$ stays below about 3\% once both the
  energy of the photon $E_\gamma$ and its opening angle $\theta$ with
  respect to the $\tau$ direction are required to be not too small.
  Those cuts are needed because of an infrared and a collinear
  divergence for $E_\gamma\to 0$ and $\theta\to 0$, respectively,
  which would be canceled by the virtual 3-loop correction to the
  2-body decay channel~\cite{Freitas:2009xxx}.}

It is interesting to note that the $\stauR\tau\axino$
vertex---governed by 2-loop diagrams---is sensitive to the two large
scales $f_a$ and $M_Q$; cf.~(\ref{Eq:GammaLL}). In contrast, there
appears only the scale $f_a$ in the vertices---governed by 1-loop
diagrams---that describe the interactions of axions/axinos with
photons, gluons, and gluinos mentioned above.

\begin{figure}[t!]
\includegraphics*[width=.495\textwidth]{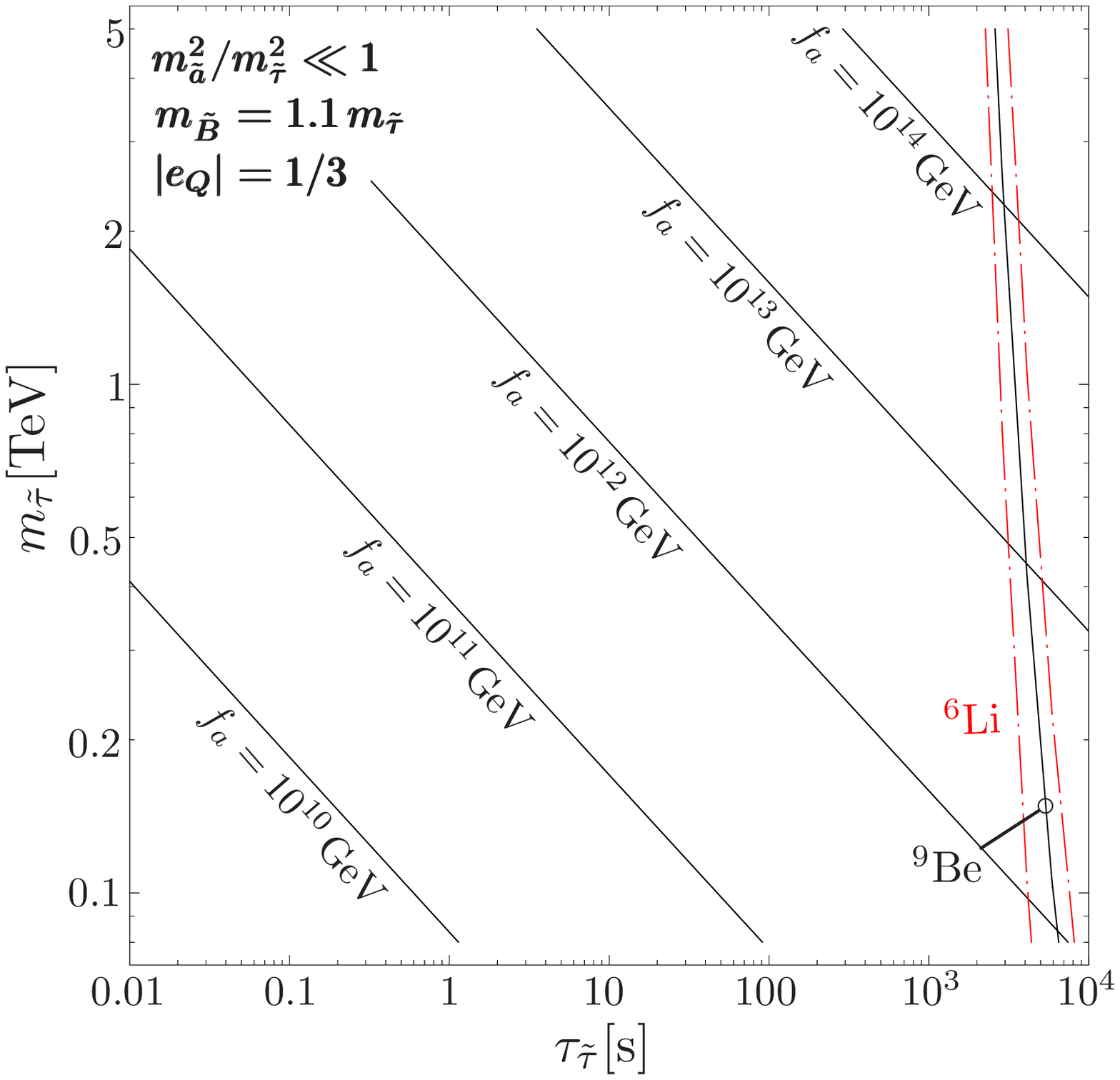}
\caption{The lifetime of the $\stauR$ NLSP,
  $1/\Gamma(\stauR\to\tau\,\axino)\approx\tau_{\stau}$ in relation to
  its mass $\mstau$ for $\maxino^2/\mstau^2\ll 1$,
  $\mbino=1.1\,\mstau$, $|e_Q|=1/3$, $y=1$, and $f_a$ values from
  $10^{10}$ to $10^{14}\,\GeV$. For a stau yield $Y_{\stau}$ given
  by~(\ref{Eq:Ystau}), $\tau_{\stau}$ values to the right of the
  nearly vertical solid and dash-dotted (red) lines are disfavored by
  the constraints~(\ref{Eq:BeNine}) and~(\ref{Eq:LiSix}) on catalyzed
  BBN (CBBN) of $^9$Be and $^6$Li,
  respectively~\cite{Pospelov:2008ta}; see Sect.~\ref{Sec:CBBNaxino}
  for details.}
\label{Fig:Stau_NLSP_Lifetime}
\end{figure}

In Fig.~\ref{Fig:Stau_NLSP_Lifetime} our result of the full leading
term for $1/\Gamma(\stauR\to\tau\,\axino)\approx \tau_{\stau}$ and its
relation to $\mstau$ is illustrated for $\maxino^2/\mstau^2\ll 1$,
$\mbino=1.1\,\mstau$, $|e_Q|=1/3$, and $y=1$. The considered $f_a$
values are between $10^{10}$ and $10^{14}\,\GeV$.

The results show that $\Gamma(\stauR\to\tau\,\axino)$ is largely
governed by the LL part~(\ref{Eq:GammaLL}). Comparing
equation~(\ref{Eq:GammaLL}) with the full
expression~\cite{Freitas:2009xxx}  (see also
Fig.~\ref{Fig:Stau_NLSP_Lifetime}), we estimate that it gives the
total width $\Gamma_{\tot}^{\stauR}$ and thereby the $\stauR$ lifetime
$\tau_{\stau} = 1/\Gamma_{\tot}^{\stauR}$ to within 10\% to maximally
15\%, depending on the values of $f_a$ and $\mstau$.

One can see that $f_a\gtrsim 10^{12}\,\GeV$ is associated with
$\tau_{\stau}> 1~\seconds$ for $\mstau\lesssim 1~\TeV$, i.e., for the
$\mstau$ range that would be accessible at the LHC.  Accordingly, BBN
constraints on axino LSP scenarios with the stau NLSP can become
important as will be discussed explicitly below.
Note that not only the LL part~(\ref{Eq:GammaLL}) but the full leading
term is strongly sensitive to the electric charge of the heavy KSVZ
fields: $\Gamma(\stauR\to\tau\,\axino)\propto e_Q^4$.
With respect to the case in Fig.~\ref{Fig:Stau_NLSP_Lifetime},
$\tau_{\stau}$ is thus reduced by a factor of 81 (16) for $|e_Q|=1$
($2/3$).
On the other hand, if $e_Q=0$, the decay of the $\stau$ NLSP will
require 4-loop diagrams involving gluons, gluinos, and ordinary
(s)quarks, which would thus lead to significantly larger lifetimes
than in Fig.~\ref{Fig:Stau_NLSP_Lifetime}.

% __________________________________________________________________
% __________________________________________________________________

Let us compare our result with the one for
$\Gamma(\stauR\to\tau\,\axino)$ that had been obtained
in~\cite{Brandenburg:2005he} with an effective theory in which the
heavy KSVZ (s)quark loop was integrated out, i.e., by using the method
described in~\cite{Covi:2002vw}. There, the logarithmic divergences
were regulated with the cut-off $f_a$, and only dominant contributions
were kept. While the dependence on the quantum numbers of the KSVZ
(s)quarks was absorbed into the constant $\CaYY$, the uncertainty
associated with this cut-off procedure was expressed in terms of a
mass scale $m$ and a factor $\xi$ in Ref.~\cite{Brandenburg:2005he}.
%
%Our 2-loop calculation allows us to eliminate this uncertainty. 
%
Our 2-loop calculation allows us to make direct connection with the
parameters of the underlying model.
In particular, we find from~(\ref{Eq:GammaLL}) that one must set
$\CaYY=6 e_Q^2$, $\xi=1$, and $m=\sqrt{2}\mstau/y$.
Assuming $y \lesssim 1$, to avoid non-perturbative heavy (s)quark
dynamics, this implies that the scale $m$ cannot be significantly
smaller than $\mstau$, which is an important result of the full 2-loop
calculation.
%
%This shows how important the full 2-loop calculation is; had one, for
%instance, set the scale $m$ to $M_Z$, a deviation of up to about 45\%
%would result at $\mstau=1~\TeV$.
%
Furthermore, the non-LL part can account, as mentioned, for up to 15\%
of the decay rate.

In the early Universe, the stau LOSP decouples as a WIMP before its
decay into the axino LSP.
The thermal relic stau abundance prior to decay then depends on
details of the SUSY model such as the mass splitting among the
lightest Standard Model superpartners~\cite{Asaka:2000zh} or the
left-right mixing of the stau LOSP~\cite{Ratz:2008qh,Pradler:2008qc}.
However, focussing on the $\stauR$ LOSP setting, we work with the
typical thermal freeze out yield described by
\begin{align}
Y_{\stau}
        \equiv
        \frac{n_{\stauR}}{s}
        = 2 Y_{\stauR^-}
        \simeq
        0.7 \times 10^{-12}
        \left(\frac{m_{\slepton}}{1~\TeV}\right),
\label{Eq:Ystau}
\end{align}
where $s$ denotes the entropy density and $n_{\stauR}$ the total
$\stauR$ number density for an equal number density of positively and
negatively charged $\stauR$'s.
This approximation~(\ref{Eq:Ystau}) agrees with the curve in Fig.~1 of
Ref.~\cite{Asaka:2000zh} derived for $\mbino=1.1\,\mstau$ and for
$\mstau$ significantly below the masses of the lighter selectron and
the lighter smuon.

\begin{figure}[t]
\includegraphics*[width=.495\textwidth]{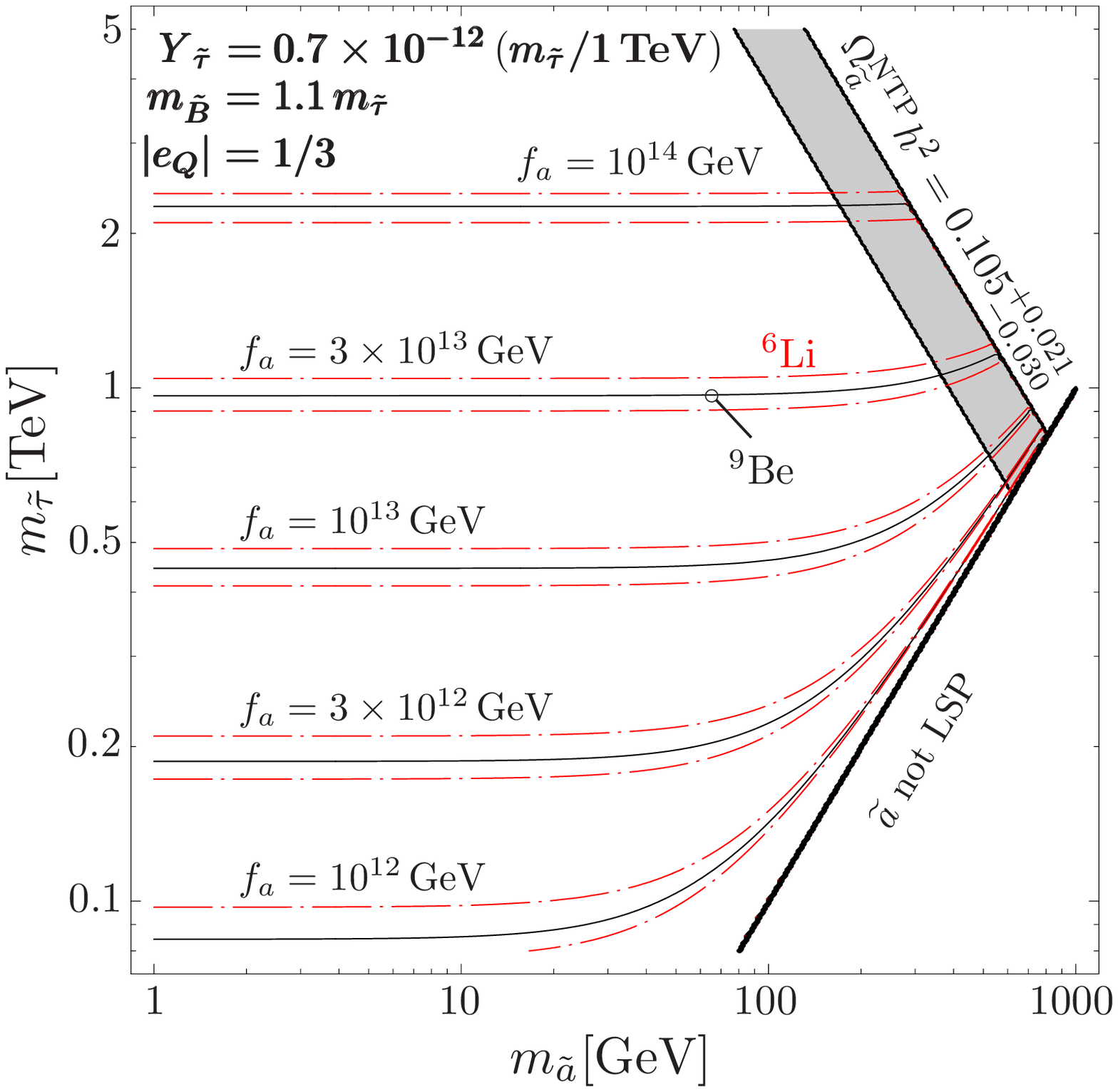}
\caption{Cosmological constraints on the masses of the $\axino$ LSP
  and the $\stauR$ NLSP for $Y_{\stau}$ given by~(\ref{Eq:Ystau}). The
  gray band indicates where $\Omega_{\axino}^{\NTP}h^2$ lies within
  the region~(\ref{Eq:OmegaDM}). Above this band,
  $\Omega_{\axino}^{\NTP}h^2>0.126$. Because of the CBBN
  reactions~(\ref{Eq:CBBNLiSix})--(\ref{Eq:ResNCapture}) becoming
  efficient, the regions below the solid and the dash-dotted (red)
  lines are disfavored by the observationally inferred limits on
  primordial $\ben$~(\ref{Eq:BeNine}) and $\Lisix$~(\ref{Eq:LiSix}),
  respectively, for $f_a$ as indicated, $\mbino=1.1\,\mstau$,
  $|e_Q|=1/3$, and $y=1$. The shown CBBN constraints thus provide
  upper limits on $f_a$ as a function of $\maxino$ and $\mstau$.
  Focussing on the $\axino$ LSP case, we do not consider the region in
  which $\maxino>\mstau$.}
\label{Fig:CBBN_Constraints}
\end{figure}

Since each stau NLSP decays into one axino LSP, the thermal relic stau
abundance leads to a non-thermally produced (NTP) axino
density~\cite{Bonometto:1993fx,Covi:1999ty,Covi:2001nw,Covi:2004rb}
\begin{equation}
        \Omega_{\axino}^{\NTP} h^2
        = 
        m_{\axino}\, Y_{\stau}\, s(T_0) h^2 / \rho_{\mathrm{c}}
        \ ,
\label{Eq:AxinoDensityNTP}
\end{equation}
where $\rho_c/[s(T_0)h^2]=3.6\times 10^{-9}\,\GeV$~\cite{Amsler:2008zz}.
For $Y_{\stau}$ given by~(\ref{Eq:Ystau}), $\Omega_{\axino}^{\NTP}h^2$
is within the nominal $3\sigma$ range~(\ref{Eq:OmegaDM}) for
$(\maxino, \mstau)$ combinations indicated by the gray band in
Fig.~\ref{Fig:CBBN_Constraints}.
%
%
% __________________________________________________________________
% __________________________________________________________________
%
%
While $\mstau$ values above this band are disfavored by
$\Omega_{\ax}^{\NTP}>\OmegaDM$, $\Omega_{\axino}^{\NTP}$ is only a
minor fraction ($\lesssim 1\%$) of $\OmegaDM$ for $\maxino\lesssim
1~\GeV$ and $\mstau\lesssim 5~\TeV$.
For $\maxino\lesssim 1~\GeV$, the $\TR$ limits shown in
Fig.~\ref{Fig:TR_Limits} will thus shift only marginally by taking
$\Omega_{\axino}^{\NTP}$ into account.

% __________________________________________________________________
\section{CBBN Constraints}
\label{Sec:CBBNaxino}
% __________________________________________________________________

The presence of negatively charged $\stauR^-$'s at cosmic times of
$t>10^3\,\seconds$ can allow for primordial $^6$Li and $^9$Be
production via the formation of $(\Hefour\,\stauR^-)$ and
$(\beetm\,\stauR^-)$ bound states. Indeed, depending on the lifetime
$\tau_{\stau}$ and the abundance $Y_{\stauR^-}=Y_{\stau}/2$, the
following catalyzed BBN (CBBN) reactions can become
efficient~\cite{Pospelov:2006sc,Pospelov:2007js,Pospelov:2008ta}%
\footnote{The large $\ben$-production cross section reported and used
  in Refs.~\cite{Pospelov:2007js,Pospelov:2008ta} has recently been
  questioned by Ref.~\cite{Kamimura:2008fx}, in which a study based on
  a four-body model is announced as work in progress to clarify the
  efficiency of $\ben$ production.}
\bea
(\Hefour\,\stauR^-)+\deuterium & \rightarrow & \Lisix + \stauR^-
\label{Eq:CBBNLiSix}
\\
\hef + (\hef\,\stauR^-) & \rightarrow & (\beetm\,\stauR^-)+\gamma
\label{Eq:RadFusion}
\\
(\beetm\,\stauR^-)+n & \rightarrow & \ben+\stauR^- \ .
\label{Eq:ResNCapture}
\eea
Observationally inferred limits on the primordial abundances of both
\lisx\ and \ben\ can thus be used to extract $\tau_{\stau}$-dependent
upper bounds on $Y_{\stauR^-}$. 
In this Letter, we adopt those bounds directly from Fig.~5 of
Ref.~\cite{Pospelov:2008ta} relying on observationally inferred limits
on the primordial fractions of
$\Lisix$~\cite{Cyburt:2002uv,Asplund:2005yt,Jedamzik:2007qk} and
$\ben$~\cite{Pospelov:2008ta} of respectively
\bea
  \Lisix/\mathrm{H}|_{\mathrm{obs}} 
  & \leq & 
  10^{-11}\!-\!10^{-10}
  \ ,
\label{Eq:LiSix}\\
  \ben/\mathrm{H}|_{\mathrm{obs}} 
  & \leq &
  2.1\times 10^{-13}
  \ .
\label{Eq:BeNine}
\eea
Confronting the $\tau_{\stau}$-dependent $Y_{\stauR^-}$ bounds
with~(\ref{Eq:Ystau}), we obtain the CBBN constraints shown in
Figs.~\ref{Fig:Stau_NLSP_Lifetime} and~\ref{Fig:CBBN_Constraints} by
the solid ($\ben$) lines and by pairs of dash-dotted ($\Lisix$, red)
lines associated, respectively, with~(\ref{Eq:BeNine}) and the range
in~(\ref{Eq:LiSix}).
The regions to the right of the corresponding lines in
Fig.~\ref{Fig:Stau_NLSP_Lifetime} and the ones below the corresponding
lines in Fig.~\ref{Fig:CBBN_Constraints} are disfavored by CBBN due to
an excess of $\ben$ and $\Lisix$ over the given limits.

In Fig.~\ref{Fig:CBBN_Constraints}, $f_a$ values from $10^{12}$ up to
$10^{14}\,\GeV$ are considered for $\mbino=1.1\,\mstau$, $|e_Q|=1/3$,
and $y=1$.
For $f_a\lesssim 10^{12}\,\GeV$ and $\maxino^2/\mstau^2\ll 1$, the
$\mstau$ values disfavored by CBBN are already excluded by the limit
$\mstau\gtrsim 80~\GeV$~\cite{Amsler:2008zz} from searches for
long-lived staus at the Large Electron Positron (LEP) collider; see
also Fig.~\ref{Fig:Stau_NLSP_Lifetime}.
Thus, for $f_a < 10^{12}\,\GeV$ and $\mstau\gtrsim 80\,\GeV$, CBBN
constraints can only be effective if $\maxino$ and $\mstau$ are
degenerate leading to a significant phase space suppression resulting
in $\tau_{\stau}>10^3\,\seconds$.
For $|e_Q|=1$, the CBBN constraints agree basically with
the contours shown in Fig.~\ref{Fig:CBBN_Constraints} but with $f_a$
values shifted upwards by one order of magnitude.

The CBBN constraints follow contours of constant $\tau_{\stau}$.
Indeed, for $\maxino^2/\mstau^2\ll 1$, the CBBN constraints also become
independent of $\maxino$.
Moreover, for given $f_a$, $\maxino$, and $\mstau$, larger values of
$\tau_{\stau}$ and thereby more restrictive CBBN constraints are
encountered at smaller values of $e_Q$, $\mbino$, or $y$.
By decreasing $\mbino$ towards $\mstau$, the CBBN constraints become
more restrictive because of both a larger $\tau_{\stau}$ and a yield
$Y_{\stau}$ that is enhanced by stau--bino coannihilation.  However,
the effect is dominated by the change in $\tau_{\stau}$ due to the relatively 
mild impact of $Y_{\tilde{\tau}^-_{\rm R}}$ on the CBBN processes in the 
relevant region; see Fig.~5 of Ref.~\cite{Pospelov:2008ta}.

Let us stress that each set of CBBN constraints in
Fig.~\ref{Fig:CBBN_Constraints}---such as the $\ben$
contours---imposes an upper limit on the PQ scale $f_a$ as a function
of $\maxino$ and $\mstau$.
Since those $f_a$ limits become only more restrictive for
$\maxino\to\mstau$, their $\maxino$-independent values at
$\maxino^2/\mstau^2\ll 1$ are conservative limits.
In the considered $\axino$ LSP case, those are relevant for studies
and searches of axions even without further insights into $\maxino$.

% __________________________________________________________________
\section{Probing \boldmath$\TR$ with BBN and at Colliders}
\label{Sec:ProbingTR}
% __________________________________________________________________
    
If the considered $\axino$ LSP scenario is realized in nature with not
too heavy Standard Model superpartners, one will be able to measure
$\mstau$ and $\mbino$ at the LHC.  Moreover, with further experimental
insights into the SUSY model, $Y_{\stau}$ can be calculated for a
standard cosmological history with $\TR$ above the temperature at
which the stau decouples from the primordial plasma.  For
concreteness, let us assume that $\mbino=1.1\,\mstau$ and that the
resulting yield agrees with~(\ref{Eq:Ystau}). The measured $\mstau$
value can then be confronted with the CBBN constraints shown in
Figs.~\ref{Fig:Stau_NLSP_Lifetime} and \ref{Fig:CBBN_Constraints}.
For $\mstau=500~\GeV$, for example, the CBBN constraints imply
$f_a\lesssim 10^{13}\,\GeV$ for $\maxino^2/\mstau^2\ll 1$,
$|e_Q|=1/3$, and $y=1$. Then $\TR\gtrsim 10^9\,\GeV$---as required by
standard thermal leptogenesis---will only be viable for
$\maxino\lesssim 1~\MeV$; cf.\ Fig.~\ref{Fig:TR_Limits}.
While $\tau_{\stau}$ is practically independent of such a small
$\maxino$, one could in principle test this $\maxino$ limit from the
kinematics of the 2-body decay
$\stauR\to\tau\axino$~\cite{Brandenburg:2005he}, i.e., from a
measurement of the energy of the emitted tau $E_\tau$,
\begin{align}
  \maxino=\sqrt{{\mstau^2}+{m_\tau^2}-2{\mstau E_\tau}} 
  \ .
\label{Eq:Axino_Mass} 
\end{align}
At present, however, this seems to be a realistic option only for
$0.1\mstau\lesssim\maxino<\mstau$ in light of the expected
experimental uncertainties. Indeed, for $\maxino\lesssim 1~\GeV$, an
experimental determination of $\maxino$ along~(\ref{Eq:Axino_Mass})
will be extremely challenging.
Nevertheless, for a given hadronic axion model (i.e., given $e_Q$ and
$y$), the CBBN constraints together with experimental insights into
$\mstau$, $\mbino$, $Y_{\stau}$, and $\OmegaDM$ imply new
$\maxino$-dependent upper limits on the reheating temperature $\TR$.%
\footnote{Reference~\cite{Choi:2007rh} also addresses ways to probe
  $\TR$ values but based on
  $\Omega_{\axino}^{\TP}+\Omega_{\axino}^{\NTP}\leq\OmegaDM$ and on
  $\Omega_{\axino}^{\NTP}$ to be inferred from collider data and
  without considering BBN constraints in the $\axino$ LSP case with a
  $\slepton$ NLSP, which are the main results of our Letter.}

\begin{figure}[t!]
\includegraphics*[width=.495\textwidth]{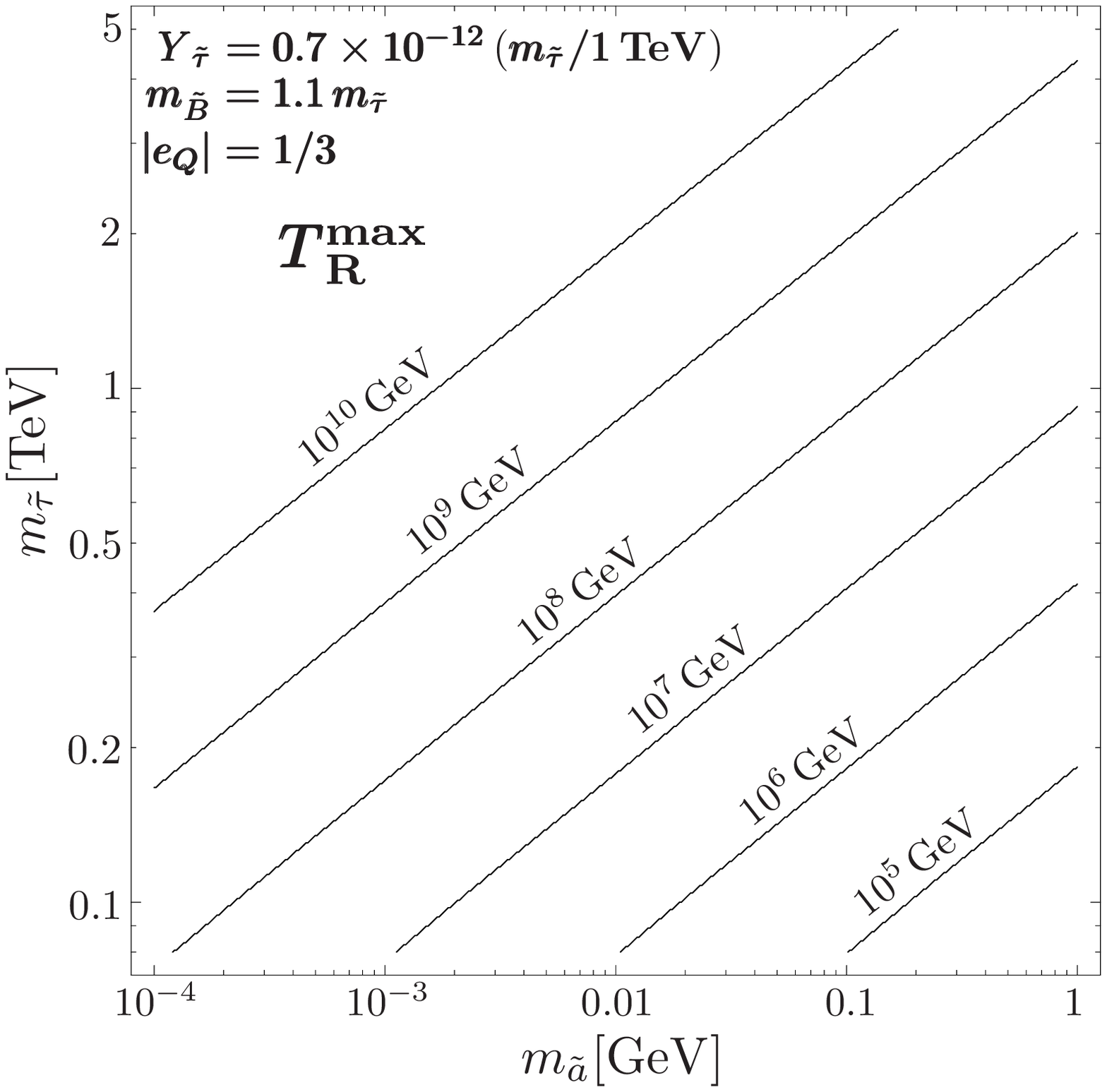}
\caption{Upper limits on the reheating temperature $\TR$ imposed by
  $\Omega_{\axino}^{\TP}h^2\leq 0.126$ and by the CBBN limit on $f_a$
  given by the upper solid line ($\ben$) in
  Fig.~\ref{Fig:CBBN_Constraints}, i.e., for $|e_Q|=1/3$,
  $\mbino=1.1\,\mstau$, $Y_{\stau}$ given by~(\ref{Eq:Ystau}), and
  $y=1$.}
\label{Fig:TRmax_CBBN}
\end{figure}

In Fig.~\ref{Fig:TRmax_CBBN}, 
%
% __________________________________________________________________
% __________________________________________________________________
%
we present upper limits on $\TR$ imposed by
$\Omega_{\axino}^{\TP}h^2\leq 0.126$ and by the $\ben$ CBBN limit on
$f_a$ given in Fig.~\ref{Fig:CBBN_Constraints}, i.e., for $|e_Q|=1/3$,
$\mbino=1.1\,\mstau$, $Y_{\stau}$ given by~(\ref{Eq:Ystau}), and
$y=1$.
The shown limits range from $\TRmax=10^{5}\,\GeV$ up to
$10^{10}\,\GeV$ (as labeled).
Once $\mstau$ is determined at colliders, this figure allows one to
infer $(\maxino,\TR)$ combinations that are disfavored by CBBN and
$\OmegaDM$.
The $\Lisix$ CBBN limits on $f_a$ are in close vicinity to the $\ben$
limit, as can be seen in Fig.~\ref{Fig:CBBN_Constraints}. Thus, we do
not show the associated $\TRmax$ lines since they agree basically with
the ones shown in Fig.~\ref{Fig:TRmax_CBBN}.
For $|e_Q|=1$, $\TRmax$ becomes less restrictive by almost exactly two
orders of magnitude. For example, the $\TRmax=10^{9}\,\GeV$ line for
$|e_Q|=1$ is in close vicinity to the $\TRmax=10^{7}\,\GeV$ line in
Fig.~\ref{Fig:TRmax_CBBN}.

The obtained upper limits on $f_a$ and $\TR$ are conservative ones.
For instance, BBN constraints from hadronic energy emitted in 4-body
decays $\stauR\to\tau\axino\quark\antiquark$ can become relevant
already for $\tau_{\stau}\gtrsim 100~\seconds$.
These additional constraints---imposed mainly by observationally
inferred limits on primordial deuterium---may imply more restrictive
$f_a$ limits than obtained here, and thereby $\TRmax$ values that are
more restrictive than the ones in Fig.~\ref{Fig:TRmax_CBBN}.
Effects of late energy injection on $\Lisix$ from CBBN have been
included in the gravitino LSP case, e.g., in
Refs.~\cite{Cyburt:2006uv,Kawasaki:2007xb,Jedamzik:2007qk,Kawasaki:2008qe}.
The resulting constraints differ only marginally from the ones
obtained without taking this effect into
account~\cite{Pradler:2007is,Pradler:2007ar,Pospelov:2008ta}.%
\footnote{At $t\lesssim 10^3\,\seconds$ when CBBN is not efficient,
  injection of energy may have a noticeable effect on the $\Lisix$
  abundance and could even allow for a solution of the $\lisv$ problem
  that is consistent with $\Lisix$ in the observationally inferred
  range~(\ref{Eq:LiSix})~\cite{Jedamzik:2004er,Jedamzik:2005dh,Cyburt:2006uv,Bailly:2008yy}.}
We expect a similar outcome for our CBBN limits and refer the study of
constraints from energy injection to a future publication.

% __________________________________________________________________
\section{Discussion}
\label{Sec:Discussion}
% __________________________________________________________________

It has already been realized in Ref.~\cite{Brandenburg:2005he} that
collider measurements of $\tau_{\stau}$, $\mstau$, and $\mbino$ will
probe the PQ scale $f_a$ in the considered axino LSP scenarios. This
is also evident from the results of our 2-loop calculation shown in
Fig.~\ref{Fig:Stau_NLSP_Lifetime} and from the associated LL
part~(\ref{Eq:GammaLL}). The $f_a$ value inferred for given $e_Q$ and
$y$ can then be used in~(\ref{Eq:AxinoDensityTP}) to extract the
$\maxino$-dependent limit $\TR^{\max}$ imposed by
$\Omega_{\axino}^{\TP}\leq\OmegaDM$; cf.\ Fig.~\ref{Fig:TR_Limits}.
However, a $\tau_{\stau}$ measurement will be challenging from the
experimental point of view. In fact, while there are proposals for
planned detectors at the International Linear Collider
(ILC)~\cite{Martyn:2006as,Martyn:2007mj}, new detector concepts may be
necessary to stop and collect long-lived $\stauone$s for an analysis
of their
decays~\cite{Hamaguchi:2006vu,Goity:1993ih,Hamaguchi:2004df,Feng:2004yi}.

The limits on $f_a$ and $\TR$ presented in
Figs.~\ref{Fig:CBBN_Constraints} and~\ref{Fig:TRmax_CBBN} do not rely
on a measurement of $\tau_{\stau}$.
They result from upper limits $\tau_{\stau}^{\max}$ imposed by the
CBBN constraints,
\begin{equation}
  \tau_{\stau}\leq\tau_{\stau}^{\max} < 10^4\,\seconds
  \ ,
\label{Eq:taumaxCBBN}
\end{equation}
which show only a very mild dependence on $\mstau$ for typical yields
such as~(\ref{Eq:Ystau}); see Fig.~\ref{Fig:Stau_NLSP_Lifetime}.
In fact, based on~(\ref{Eq:taumaxCBBN}), it is possible to derive
analytic expressions for the upper limits on $f_a$ and $\TR$ in a
conservative way.

Aiming at an instructive derivation, we work with the LL
part~(\ref{Eq:GammaLL}) which describes $\tau_{\stau}$ to within 15\%
accuracy,
\bea
        &&\!\!\!\!\!\!\!\!\!\!\!\!\!\!\!\!
        \tau_{\stau} 
        \approx 
        \tau_{\stau\,\LL}
        \equiv
        \Gamma(\stauR\to\tau\axino)_{\LL}^{-1}
\label{Eq:TauStauLL}
\\
        &&\!\!\!\!\!\!\!\!\!\!\!\!
        \gtrsim
        \frac{128\pi^5\cos^8 \theta_{W}}{81\,\alpha^4 e_Q^4}
        \,
        \frac{f_a^2}{\mstau\,\mbino^2}
        \,
        \ln^{-2} \left(\frac{y f_a}{\sqrt{2} \mstau}\right)
\label{Eq:TauStauSmallMa}
\\
        &&\!\!\!\!\!\!\!\!\!\!\!\!
        \gtrsim
        3.78\times 10^3~\seconds
	\left(\frac{1/3}{e_Q}\right)^4
        \left(\frac{f_a}{10^{12}\,\GeV}\right)^{\!\!2}\!  \nonumber \\
        &&\quad
        \times
        \left(\frac{100~\GeV}{\mstau}\right)\!
        \left(\frac{100~\GeV}{\mbino}\right)^{\!\!2}
\!\!,
\label{Eq:TauStauMIN}
\eea
where (\ref{Eq:TauStauSmallMa}) underestimates $\tau_{\stau\,\LL}$ by
at most 2\% (15\%) for
$\maxino\lesssim 0.1\mstau$ ($\maxino\lesssim 0.25\mstau$).
Focussing on the collider-friendly region $\mstau\lesssim 1~\TeV$,
$f_a\lesssim 3\times10^{13}\,\GeV$ is imposed by CBBN for $|e_Q|=1/3$
and $y=1$.
Based on this and on the LEP bound $\mstau\gtrsim 80~\GeV$,
$\ln(y f_a/\sqrt{2}\mstau)\lesssim 26.3$
is used to get from~(\ref{Eq:TauStauSmallMa})
to~(\ref{Eq:TauStauMIN}). Accordingly, $\tau_{\stau\,\LL}$ can be
underestimated by~(\ref{Eq:TauStauMIN}) by a factor of $\Order(1)$ at
$f_a\ll 3\times 10^{13}\,\GeV$ and/or $80~\GeV\ll\mstau\lesssim
1~\TeV$.
Nevertheless, (\ref{Eq:TauStauMIN}) allows us to translate the
constraint~(\ref{Eq:taumaxCBBN}) in a conservative way into the
following upper limit:
\bea
        f_a
        &\lesssim& 
        1.63 \times 10^{12}\,\GeV 
	\left(\frac{e_Q}{1/3}\right)^2
        \left(\frac{\tau_{\stau}^{\max}}{10^4\,\seconds}\right)^{1/2}
\nonumber\\
        &&
        \times 
        \left(\frac{\mstau}{100~\GeV}\right)^{1/2}
        \left(\frac{\mbino}{100~\GeV}\right)      
        \equiv
        f_a^{\max} 
\ . 
\label{Eq:famaxANAL}
\eea
A comparison with the numerically obtained $\ben$ limits at
$\maxino^2/\mstau^2\ll 1$ shows a good overall agreement for
$\tau_{\stau}^{\max}\approx 5\times 10^3\,\seconds$. The associated
analytical expression however is less restrictive (i.e., more
conservative) than the numerically obtained limits towards larger
$\mstau$. In fact, there the actual $\tau_{\stau}^{\max}$ value
imposed by CBBN becomes more restrictive as can be seen in
Fig.~\ref{Fig:Stau_NLSP_Lifetime}.

Let us now turn to $\TR$ on which a conservative limit
\begin{equation}
        \TR
        \lesssim
        1.7\times 10^6\,\GeV
        \left(\frac{\OmegaDM h^2}{0.1}\right)
        \left(\frac{f_a}{10^{11}\,\GeV}\right)^{\! 2}\!\!
        \bigg(\frac{0.1~\MeV}{m_{\ax}}\bigg)
\label{Eq:TRmaxANAL}
\end{equation}
is imposed by
\bea
        &&\!\!\!\!\!\!\!\!\!\!\!\!\!\!\!\!
        \OmegaDM h^2
        \geq
        \Omega_{\axino}^{\TP}h^2
\\
        &&\!\!\!\!\!\!\!\!\!\!\!\!
        \gtrsim
        0.6
        \left(\frac{10^{11}\,\GeV}{f_a}\right)^{\! 2}\!\!
        \bigg(\frac{m_{\ax}}{0.1~\MeV}\bigg)\!
        \left(\frac{T_R}{10^7\,\GeV}\right)
\ .
\label{Eq:OmegaAxinoMIN}
\eea
Here the constant ``conservative'' prefactor $0.6$ accounts for the
$\TR$-dependent prefactor in~(\ref{Eq:AxinoDensityTP}), which stays in
the range
$0.6<5.5\,g_\mathrm{s}^6(\TR)\ln[1.211/g_\mathrm{s}(\TR)]<1.06$
for $10^4\,\GeV\leq\TR\leq 10^{12}\,\GeV$ if the MSSM 1-loop
renormalization group running of $g_s$ is considered.
Using the upper limit~(\ref{Eq:famaxANAL}) in~(\ref{Eq:TRmaxANAL}),
one arrives immediately at an analytic expression for the CBBN-imposed
limit,
\bea
        \TR 
        &\lesssim&
        4.4\times 10^8\,\GeV
	\left(\frac{e_Q}{1/3}\right)^4
        \left(\frac{\OmegaDM h^2}{0.1}\right)
        \bigg(\frac{0.1~\MeV}{m_{\ax}}\bigg)
\nonumber\\
        &&
        \times 
        \left(\frac{\tau_{\stau}^{\max}}{10^4\,\seconds}\right)
        \left(\frac{\mstau}{100~\GeV}\right)
        \left(\frac{\mbino}{100~\GeV}\right)^{2}      
        \equiv
        \TR^{\max} 
\ , 
\nonumber\\
\label{Eq:TRmaxCBBN}
\eea
which is conservative. For $\tau_{\stau}^{\max}\approx 5\times
10^3\,\seconds$, we find again a good overall agreement with the
limits obtained numerically. However, as expected from its derivation,
the associated analytic expression can be by a factor of $\Order(1)$
less restrictive than the numerical results shown in
Fig.~\ref{Fig:TRmax_CBBN}.

Since $\tau_{\stau}$ depends on the ratio $f_a/e_Q^2$, the
limits~(\ref{Eq:famaxANAL}) and (\ref{Eq:TRmaxCBBN}) depend on $e_Q$
and thus on the specific axion model.  It would therefore be
particularly valuable to discover the axion and its mass since the
relation between $m_a$ and $f_a$ does not depend on $e_Q$;
in the models given by~(\ref{Eq:SuperpotentialPQ}) and
Table~\ref{Tab:Q_quantum_numbers},
$m_a=[\sqrt{z}/(1+z)]\,f_{\pi}m_{\pi}/f_a$ with $f_{\pi}\approx
92~\MeV$ and $m_{\pi}=135~\MeV$.
If $f_a$ can thus be determined, $\TR^{\max}$ would be given
by~(\ref{Eq:TRmaxANAL}) directly. In addition, one could find $e_Q$ in
a $\tau_{\stau}$ measurement or derive a lower limit on it from the
CBBN constraints~(\ref{Eq:taumaxCBBN}).

In this respect we note that most axion searches probe the
axion-photon-coupling $\gagg=\alpha C_{a\gamma\gamma}/(2\pi f_a)$ in
certain ranges of the axion mass $m_a$; cf.\ \cite{Steffen:2008qp} and
references therein. In the models considered, $C_{a\gamma\gamma}$ is
given by~(\ref{Eq:Cagammagamma}) so that $\gagg$ does also depend on
$f_a$ and $e_Q$~\cite{Kim:1998va}. An axion discovery at an
($m_a$,\,$\gagg$) combination would thus be associated with an
($f_a$,\,$e_Q$) combination in the considered models.
The $e_Q$ value from axion searches could then be compared to the one
inferred from a $\tau_{\stau}$ measurement at colliders or, if this is
not possible, to its lower limit imposed by CBBN.

The region in which the presented BBN constraints are expected to
become relevant is explored by the axion dark matter experiment (ADMX)
which searches for resonant conversion of dark matter axions into
photons in a microwave cavity~\cite{Duffy:2006aa,Carosi:2007uc}.
Axion searches of this type are sensitive to $\gagg$ only in the
combination $\gagg^2 \rho_a$, where $\rho_a$ denotes the local halo
density of axions.
If axinos provide the dominant component of cold dark matter, $\rho_a$
can be very small so that no signals will appear at the expected
$\gagg$ values. 
An axion signal in such a direct search would in turn imply a sizeable
axion density, $\Omega_a\sim\OmegaDM$, and thereby a restrictive $\TR$
limit in the considered $\maxino$ range, $\maxino\gtrsim 0.1~\MeV$,
given by~(\ref{Eq:TRmaxANAL}) or~(\ref{Eq:TRmaxCBBN}) with
$\OmegaDM\to\OmegaDM-\Omega_a$.
Alternatively, evidence for solar axions could appear in the Tokyo
Axion Helioscope or the CERN Axion Solar Telescope
(CAST)~\cite{Andriamonje:2007ew,Ruz:2008zz,Minowa:2008uj}. This would
imply $\Omega_a\ll\OmegaDM$, $f_a\lesssim 10^9\,\GeV$ and thus $\TR\ll
10^6\,\GeV$ in the considered axino cold dark matter scenarios; cf.\ 
(\ref{Eq:TRmaxANAL}) with $\maxino\gtrsim 0.1~\MeV$.
Here the CBBN constraints will be relevant only in the exceptional
cases with $e_Q\to 0$ and/or $\maxino\to\mstau$.

% __________________________________________________________________
\section{Summary and Conclusions}
\label{Sec:Conclusion}
% __________________________________________________________________

We have explored BBN constraints in axino cold dark matter scenarios
with a long-lived charged slepton $\slepton$.
Calculating the lifetime $\tau_{\slepton}$, which is governed by
2-loop diagrams in hadronic axion models, we find that $\slepton$ can
be sufficiently long lived to allow for an efficient catalysis of
$\lisxm$ and $\ben$ via bound-state formation
with primordial nuclei.
Observationally inferred abundances of $\lisxm$ and $\ben$ thus impose
upper limits on $\tau_{\slepton}$ for typical thermal relic abundances
of the long-lived $\slepton$. 
These limits have allowed us to derive upper limits on the PQ scale
$f_a$ that depend mainly on the masses of the slepton, $\mslepton$,
and the lightest neutralino, $m_{\neutralino}$, and on the electric
charge of the heavy (s)quarks $e_Q$.
The obtained $f_a$ constraints imply new upper limits on the reheating
temperature $\TR$ since $f_a$ governs not only $\tau_{\slepton}$ but
also the efficiency of thermal axino production and thereby the $\TR$
constraints imposed by $\Omega_{\axino}^{\TP}\leq\OmegaDM$.
We have presented both numerical results and analytical approximations
for those new BBN-imposed limits and have discussed their dependence
on $\maxino$, $\mslepton$, $m_{\neutralino}$, and $e_Q$.
For example, for $\mslepton=500~\GeV$,
$m_{\neutralino}=1.1~\mslepton$, and $|e_Q|=1/3$, we find $f_a\lesssim
10^{13}\,\GeV$ and that $\TR\gtrsim 10^9\,\GeV$ is viable only for
$\maxino\lesssim 1~\MeV$.

We have addressed the extent to which the BBN-imposed limits on $f_a$
and $\TR$ can be probed experimentally
if the considered axino LSP scenario is realized.
With not too heavy Standard Model superpartners, LHC experiments will
allow us to measure $\mslepton$ and $m_{\neutralino}$ and to infer the
thermal relic $\slepton$ abundance prior to decay under the assumption
of a standard cosmological history.
With the ILC and/or new detector concepts, even a measurement of
$\tau_{\slepton}$ is conceivable, and our
$\tau_{\slepton}$ result shows that this could give insights into
$f_a/e_Q^2$.
A determination of $\maxino$ however seems possible only for
relatively heavy axinos $0.1\mslepton\lesssim\maxino<\mslepton$ and
hopeless for $\maxino^2/\mslepton^2\ll 1$~\cite{Brandenburg:2005he}.
Moreover, insights into $e_Q$---or, more generally, into the axion
model---seem to require not only an axion discovery but a
determination of its mass $m_a$ (and thereby of $f_a$) in axion search
experiments.

A simple form of the superpotential has been considered that is
generic for SUSY hadronic axion models in which the axion multiplet
interacts with the MSSM multiplets through loops of heavy 
(s)quarks.
While we have explored the case with a minimum number of
SU(2)$_\Weak$-singlet KSVZ multiplets and with $\slepton$ being a purely
right-handed stau $\stauR$, our study can be generalized to more
complicated settings in a straightforward way.

Without specifying the SUSY breaking mechanism or other details of the
PQ sector, we have assumed saxion effects to be negligible and a
spectrum with the $\axino$ LSP and the $\slepton$ NLSP.
Our results depend crucially on these assumptions.
In situations in which the saxion dominates the energy density before
its decay, the entropy per comoving volume can be enhanced by a factor
$\Delta>1$. If this additional entropy production takes place before
$\slepton$ decoupling, the BBN constraint on $f_a$ will not be
affected but the thermally produced axino density can be diluted so
that $\Omega_{\axino}^{\TP}\to\Omega_{\axino}^{\TP}/\Delta$ and
$\TR^{\max}\to\Delta\TR^{\max}$. If entropy increases by a large
factor of $\Delta>10^3$ after $\slepton$ decoupling and before BBN,
the $\slepton$ abundance can be diluted such that catalyzed BBN (CBBN)
of $^6$Li and $^9$Be cannot become efficient. Then the CBBN-imposed
constraints on $f_a$ and $\TR$ would not exist. Nevertheless,
$\Omega_{\axino}^{\TP}\to\Omega_{\axino}^{\TP}/\Delta$ so that the
$\OmegaDM$-imposed limit on $\TR$ would be relaxed by a factor of
$\Delta$. However note that the baryon asymmetry would also be diluted
by a factor of $\Delta$ and therefore a larger asymmetry would be
needed before its dilution;
see Ref.~\cite{Pradler:2006hh} for a related discussion in the
$\gravitino$ LSP case.

The cosmological constraints presented in this work can also be
affected by the presence of the gravitino $\tilde{G}$ even for a
standard thermal history. Its mass $m_{\tilde{G}}$---which depends on
the SUSY breaking mechanism and the SUSY breaking scale---governs the
strength of its interactions.  The gravitino can be produced thermally
in the early Universe, with the resulting abundance depending on
$m_{\tilde{G}}$ and $T_R$~\cite{Bolz:2000fu,Pradler:2006qh}. In the
scenario studied in this Letter, $m_{\tilde{a}} < m_{\tilde{l}_1} <
m_{\tilde{G}}$, the heavy gravitino is typically long-lived and its
decays may affect BBN.  Thereby additional constraints on $T_R$ can be
incurred~\cite{Kawasaki:2008qe,Kohri:2005wn}.

If $m_{\tilde{G}} < m_{\tilde{l}_1}$ and $\Gamma(\tilde{l}_1 \to l
\tilde{a}) \ll \Gamma(\tilde{l}_1 \to l \tilde{G})$, $\tau_{\slepton}$
is governed by $\slepton\to\lepton\gravitino$. Then our $f_a$ limit
can be evaded while the CBBN constraints discussed
in~\cite{Cyburt:2006uv,Steffen:2006wx,Pradler:2006hh,Pradler:2007is,Pradler:2007ar,Steffen:2008bt,Kawasaki:2008qe,Pospelov:2008ta}
and their implications for thermally produced gravitino abundance
become relevant.
On the other hand, if $\Gamma(\tilde{l}_1 \to l \tilde{a}) \gg
\Gamma(\tilde{l}_1 \to l \tilde{G})$, the CBBN limits discussed in
this Letter also apply. However, the gravitinos lead to an increase of
the LSP density, thus leading to more restrictive $T_R$ limits. In
this case our results remain as conservative upper limits.

Our investigations show that for the interesting case of new
long-lived charged particles, BBN constraints play an important role
and can be used to restrict the models considerably.
These constraints will become particularly important if such particles
are produced and detected at the upcoming LHC experiments.

%\medskip

\begin{acknowledgments}
  We are grateful to Koichi Hamaguchi, Kazunori Nakayama, Josef
  Pradler, Sabine Schilling, and Fuminobu Takahashi for valuable
  discussions.
  NT would like to thank the Max Planck Institute for Physics for
  their kind hospitality during parts of this work.
  DW would like to thank SLAC for their kind hospitality during parts
  of this work.
  This research was partially supported by the Cluster of Excellence
  `Origin and Structure of the Universe' and by the Swiss National
  Science Foundation (SNF) under contract 20-116756/2.
\end{acknowledgments}
%
% __________________________________________________________________

% __________________________________________________________________
%
%
\end{document}